\documentclass[12pt,aps,prl,onecolumn,showpacs,superscriptaddress,amsmath,amssymb]{revtex4-2}

\usepackage{setspace}

\usepackage{graphicx}
\usepackage{dcolumn}
\usepackage{bm}
\usepackage{amsmath}
\usepackage{amssymb}
\usepackage{indentfirst}
\usepackage{booktabs}
\usepackage{multirow}
\usepackage{colortbl}
\usepackage{epsfig}
\usepackage[shortcuts]{extdash}
\usepackage[colorlinks=true, linkcolor=red, citecolor=blue, urlcolor=red]{hyperref}

\begin{document}

\onehalfspacing

\title{Fractional Quantum Multiferroics from Coupling of Fractional Quantum Ferroelectricity and Altermagnetism}

\author{M. Q. Dong}
\email{These authors contributed equally to this work.}
\affiliation{Key Laboratory of Computational Physical Sciences (Ministry of Education), Institute of Computational Physical Sciences, State Key Laboratory of Surface Physics, and Department of Physics, Fudan University, Shanghai 200433, China}
\affiliation{State Key Laboratory for Mechanical Behavior of Materials, School of Materials Science and Engineering, Xi’an Jiaotong University, Xi’an, Shanxi 710049, China}

\author{B. Liu}
\email{These authors contributed equally to this work.}
\affiliation{Key Laboratory of Computational Physical Sciences (Ministry of Education), Institute of Computational Physical Sciences, State Key Laboratory of Surface Physics, and Department of Physics, Fudan University, Shanghai 200433, China}
\affiliation{State Key Laboratory for Mechanical Behavior of Materials, School of Materials Science and Engineering, Xi’an Jiaotong University, Xi’an, Shanxi 710049, China}

\author{Z. H. Dai}
\affiliation{Key Laboratory of Computational Physical Sciences (Ministry of Education), Institute of Computational Physical Sciences, State Key Laboratory of Surface Physics, and Department of Physics, Fudan University, Shanghai 200433, China}

\author{Zhi-Xin Guo}
\email{zxguo08@xjtu.edu.cn}
\affiliation{State Key Laboratory for Mechanical Behavior of Materials, School of Materials Science and Engineering, Xi’an Jiaotong University, Xi’an, Shanxi 710049, China}

\author{Hongjun Xiang}
\affiliation{Key Laboratory of Computational Physical Sciences (Ministry of Education), Institute of Computational Physical Sciences, State Key Laboratory of Surface Physics, and Department of Physics, Fudan University, Shanghai 200433, China}

\author{Xin-Gao Gong}
\email{xggong@fudan.edu.cn}
\affiliation{Key Laboratory of Computational Physical Sciences (Ministry of Education), Institute of Computational Physical Sciences, State Key Laboratory of Surface Physics, and Department of Physics, Fudan University, Shanghai 200433, China}

\date{\today}

\begin{abstract}

Multiferroics, which combine ferroelectric and magnetic order, offer a transformative platform for next-generation electronic devices. However, the intrinsic competition between the mechanisms driving ferroelectricity and magnetism in single-phase materials severely limits their performance, typically resulting in weak magnetoelectric coupling at room temperature. Here, we propose a solution to this long-standing challenge through the novel concept of fractional quantum multiferroics (FQMF), where strong magnetoelectric coupling is naturally realized by coupling fractional quantum ferroelectricity (FQFE) with altermagnetism (AM). Symmetry analysis shows that reversing the FQFE polarization necessarily inverts the AM spin splitting under parity–time ($\mathcal{PT}$) or time-reversal ($\mathcal{T}\tau$) operations. A minimal tight-binding model reproduces this effect, demonstrating electrically driven spin control without rotating the Néel vector. First-principles calculations further identify a broad family of candidate materials in two and three dimensions including bulk MnTe, Cr$_2$S$_3$, Mn$_4$Bi$_3$NO$_{15}$ and two-dimensional AB$_2$ bilayers such as MnX$_2$ (X=Cl, Br, I), CoCl$_2$, CoBr$_2$, and FeI$_2$. Notably, MnTe exhibits a high Néel temperature ($\sim$300 K) and a large electrically switchable spin splitting ($\sim$0.8 eV), demonstrating room-temperature magnetoelectric performance that surpasses that of conventional multiferroics. To further showcase the technological potential, we propose an electric-field-controlled FQMF tunnel junction based on MnTe that achieves tunneling magnetoresistance exceeding 300\%. This work establishes FQMF as a distinct and promising route to achieving room-temperature strong magnetoelectric coupling, opening a new avenue for voltage-controlled spintronics.

\end{abstract}

\maketitle


\section{\label{sec:introduction}$\text{Introduction}$}

Multiferroic materials, which simultaneously exhibit ferroelectric and magnetic ordering \cite{Schmid01011994, PhysRevLett.125.037203, van2004origin, kimura2003magnetic, song2022evidence, ponet2022topologically}, hold significant promise for next-generation memory devices, sensors, and spintronic applications \cite{ramesh2007multiferroics, fiebig2016evolution, schmid1994multi}. However, in single-phase systems, ferroelectricity and magnetism are intrinsically competitive: ferroelectricity typically relies on lattice distortions and requires polar point group symmetry, while magnetism often originates from partially filled $d$ orbitals that inherently suppress such distortions \cite{khomskii2006multiferroics, hill2000there}. This fundamental incompatibility hinders the coexistence of robust ferroelectricity and strong magnetism, resulting in a scarcity of room-temperature multiferroics with strong magnetoelectric coupling.

Recent developments in fractional quantum ferroelectricity (FQFE) offer a promising path to circumvent this limitation \cite{PhysRevLett.134.016801, ji2024fractional, trhd-kxm1}. FQFE enables switchable polarization in nonpolar point groups, without the need for conventional lattice distortions. Integrating magnetism into FQFE systems is predicted to give rise to a novel class of multiferroics [termed fractional quantum multiferroics (FQMF) in this study] \cite{PhysRevLett.134.016801, ji2024fractional}. In particular, by operating within nonpolar symmetries, FQMF can potentially overcome the traditional incompatibility between strong ferroelectricity and strong magnetism, thus providing a route to achieving room-temperature magnetoelectric coupling. However, the realization of intrinsic magnetoelectric coupling in FQMF systems remains an open question.

In this context, Altermagnetism (AM) emerges as an ideal magnetic partner for FQFE in forming FQMF \cite{PhysRevX.12.031042, PhysRevX.12.040501, PhysRevX.12.021016}. AM exhibits momentum-dependent spin splitting [$\Delta E_n (\mathbf{k})=E_n^{\uparrow} (\mathbf{k})-E_n^{\downarrow} (\mathbf{k})$] while retaining the advantaged of antiferromagnetic, including high stability and the absence of stray magnetic fields \cite{jungwirth2016antiferromagnetic, baltz2018antiferromagnetic}. Several groups have demonstrated that multiferroics composed of AM and conventional ferroelectricity can achieve symmetry-protected magnetoelectric coupling \cite{PhysRevLett.134.106802, altermagfer, sun2025proposing, zhu2025two}. However, these systems are still limited by the inherent competition between polar order and magnetism in polar point groups, which restricts the magnitude of spin splitting to values typically below 50 meV \cite{PhysRevLett.134.106802, altermagfer, sun2025proposing, zhu2025two}.

In this Letter, we demonstrate that FQMF with strong intrinsic magnetoelectric coupling arises naturally from the synergy between FQFE and AM. Through symmetry analysis, we show that reversing the FQFE polarization inevitably inverts the AM spin texture under combined parity–time ($\mathcal{PT}$) or time-reversal ($\mathcal{T}\tau$) operations, where $\tau$ denotes a fractional lattice translation. This mechanism is captured by a minimal tight-binding model, which confirms the electrical control of spin splitting without altering the Néel vector. Using first-principles calculations, we identify a broad family of two- (2D) and three-dimensional (3D) candidate materials—all residing in nonpolar point groups—that host this FQMF order. As a proof of concept, we design an electric-field-controlled fractional quantum multiferroic tunnel junction (FQMFTJ) that achieves a tunneling magnetoresistance exceeding 300\%, highlighting the potential of FQMF systems for high-performance spintronic applications.

\section{\label{sec:mechanism}$\text{Physical mechanism}$}

According to the modern theory of polarization, atomic displacements induce ferroelectric polarization \cite{modernpolar, spaldin2012beginner}. In FQFE systems, fractional lattice displacements occur when ion(s) move from high-symmetry points or lines to another equivalent position, maintaining the energy and symmetry of both the initial ($L_1$) and final states ($L_2$) \cite{PhysRevLett.134.016801, ji2024fractional}. This enables the emergence of ferroelectric polarization in nonpolar point groups. Introducing magnetism into a FQFE system yields a new phase that simultaneously hosts ferroelectric and magnetic order, i.e., FQMF. Device functionality requires direct coupling between these order parameters. Whereas conventional coupling relies on spin–orbit coupling (SOC) \cite{song2022evidence}, AM provides an alternative route that locks spin splitting to the crystal structure even without spin–orbit coupling. Once AM is embedded into FQFE structure, symmetry alone may enforce electrically controlled magnetism. Specifically, if the $L_1$ and $L_2$ structures related by $\mathcal{T}\tau$ or by the $\mathcal{PT}$ operation, or equivalently by $\tau\mathcal{PT}$ operation, the ferroelectric switching necessarily reverses $\Delta E_n (\mathbf{k})$ without reorienting the Néel vector. Thus, the coupling of FQFE and AM yields an electric-field-switchable spin splitting.

\begin{figure}
	\centering
	\includegraphics[width=\linewidth]{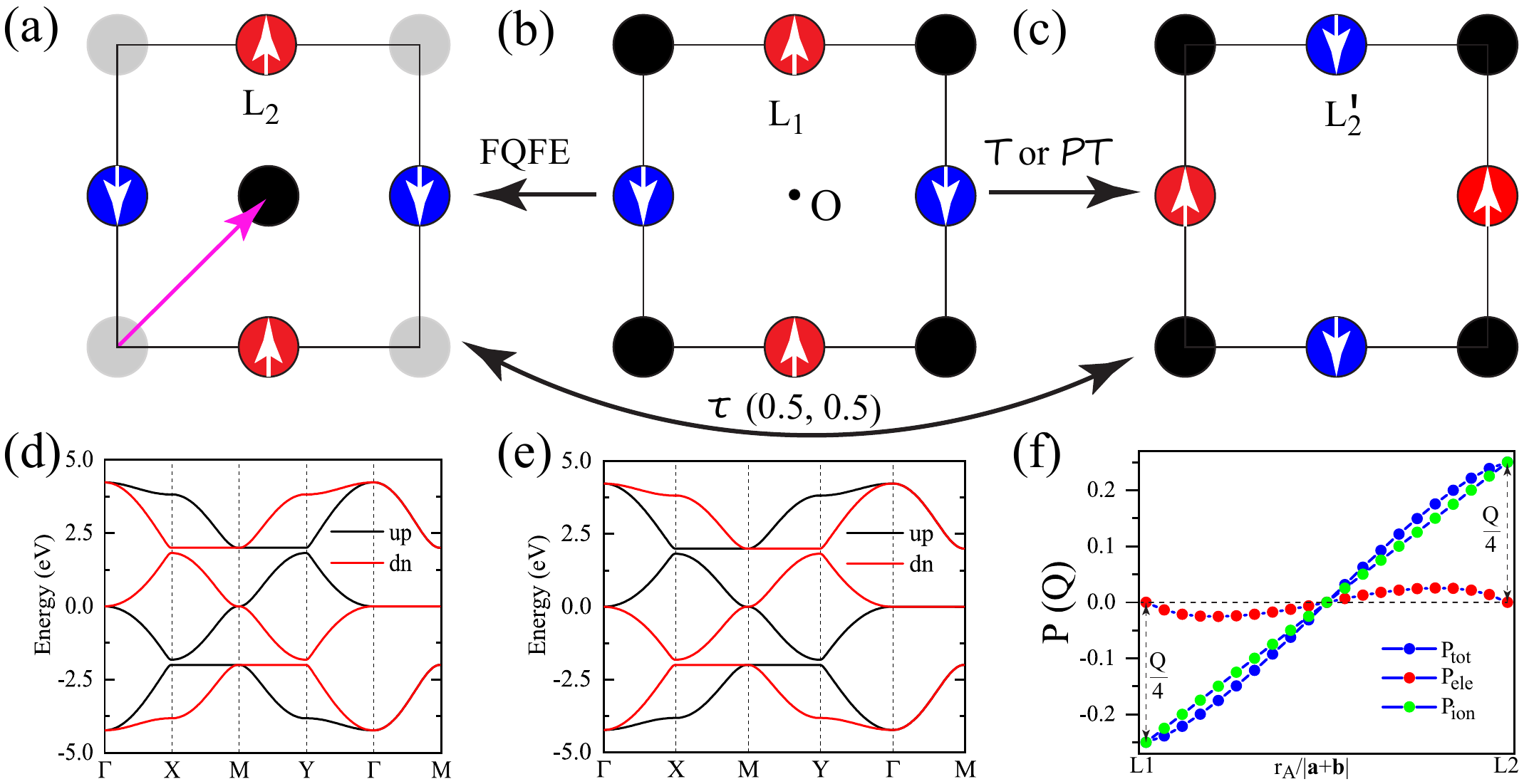}
	\caption{\label{fig:model}Schematic illustration of FQMF in monolayer AB$_2$. (a)–(c) Atomic structures for $L_2$, $L_1$, and $L_2^\prime$, respectively. The A atom (black sphere) is non-magnetic, while B atoms (red and blue spheres) carry opposite magnetic moments. The displacement of the A atom is indicated by the pink arrow, and the white arrow indicates the spin of the B atoms. In panel (b), the point O marks the origin of the inverse operation. It is important to note that although (b) exhibits inversion symmetry, this is not a requirement for our proposed FQMF. A specific example can be seen in MnBr$_2$ [Fig. (\ref{fig:mnbr2})]. (d) and (e) Band structures for $L_2$ and $L_1$, respectively. The coordinates of the high-symmetry points are provided in Sec. SIV of the Supplemental Material \cite{supplementary}. (f) Polarization difference from $L_1$ to $L_2$. $\boldsymbol{Q}$ denotes the polarization quantum \cite{modernpolar}; $P_{tot}$, $P_{ele}$, and $P_{ion}$ are the total, electronic, and ionic contributions, respectively.}
\end{figure}

We first illustrate the switching mechanism of FQMF within a phenomenological model. Figures \ref{fig:model}(a) and \ref{fig:model}(b) show the switching process in the 2D monolayer candidate AB$_2$: panel (a) depicts the final state $L_2$, and panel (b) the initial state $L_1$. In this process, the nonmagnetic atom (denoted as A atom in AB$_2$ monolayer) translates from the position (0, 0) to (0.5, 0.5). On the other hand, the $L_1$ and $L_2$ states are also related by the operations $\mathcal{T}\tau$ or $\tau\mathcal{PT}$. To make this explicit, we translate the $L_2$ state of Fig. \ref{fig:model}(a) by $\tau^{-1}=(-0.5, -0.5)$ to obtain the auxiliary state $L_2^\prime$ shown in Fig. \ref{fig:model}(c). $L_2^\prime$ is then mapped onto $L_1$ by either $\mathcal{T}$ or $\mathcal{PT}$ operation, so that $L_2 = \tau L_2^\prime=\tau\mathcal{PT}L_1$ or $L_2 = \tau L_2^\prime=\tau\mathcal{T}L_1$. For nonrelativistic collinear magnets \cite{PhysRevX.12.031042, PhysRevX.12.040501, PhysRevX.12.021016}, we have
\begin{eqnarray}
	\tau\mathcal{PT}\epsilon_n(\mathbf{k},\mathbf{s}) = \epsilon_n(\mathbf{k},-\mathbf{s})
	\label{eq:ek1}
\end{eqnarray}
and
\begin{eqnarray}
	\tau\mathcal{T}\epsilon_n(\mathbf{k},\mathbf{s}) = \epsilon_n(-\mathbf{k},-\mathbf{s}) = \epsilon_n(\mathbf{k},-\mathbf{s})
	\label{eq:ek2}
\end{eqnarray}
showing that $L_1$ and $L_2$ have opposite spin polarization band structures. This means that in FQMF systems, atomic displacements can simultaneously induce (or reverse) ferroelectric polarization and invert the spin polarization $\Delta E_n (\mathbf{k})$, while leaving the Néel vector unchanged. 

\section{\label{sec:model}$\text{Tight-binding model}$}
We further construct an effective tight-binding (TB) model on a 2D square lattice that hosts AM order, as shown in Figs. \ref{fig:model}(a)–\ref{fig:model}(c), to elucidate the microscopic origin of FQMF. The electronic Hamiltonian can be written as
\begin{eqnarray}
	H=-\sum_{\langle ij\rangle, s}{t_{ij}\left( \boldsymbol{r}_A \right) c_{i,s}^{\dagger}c_{j,s}}-J\sum_{i,s,s^\prime}{\mathrm{M}_ic_{i,s}^{\dagger}\mathbf{\sigma }_{ss^\prime}^{z}c_{i,s^\prime}}-\mu \sum_{i,s}{c_{i,s}^{\dagger}c_{i,s}}.
	\label{eq:model}
\end{eqnarray}
Here, $c_{i,s}^{\dagger}$ and $c_{i,s}$ are fermionic creation and annihilation operators with spin $s\in (\uparrow,\downarrow)$ at site $i$. The notation $\langle ij\rangle$ denotes neighbor pairs where distance $r_{ij}<a$, $a$ is the lattice constant. The hopping parameter $t_{ij}$  depends on the overlap of atomic wave functions and thus varies with the position of atom A, coupling it to the ferroelectric polarization. To illustrate how FQFE polarization varies with ionic displacement while preserving structural symmetry, we choose
\begin{eqnarray}
	t_{ij} = 
	\begin{cases}
		4.5\left[1-\sin(\frac{\pi r_{ij}}{2a})\right], & r_{ij} <a. \\
		0, & r_{ij} \ge a.
	\end{cases}
	\label{eq:tij}
\end{eqnarray}
The $J$ term describes the on-site exchange interaction between the local spin $M_i$ and itinerant electron spin $s$. The chemical potential $\mu$ controls the electron filling of the system (see Supplemental Material \cite{supplementary} Sec. I for details).
\nocite{modernpolar,PhysRevB.54.11169, PhysRevB.47.558, PhysRevLett.77.3865, PhysRevB.57.1505, PhysRevB.59.1758, PhysRevB.50.17953, PhysRevB.47.1651, resta1992theory, RevModPhys.66.899, PhysRevB.63.245407, PhysRevB.65.165401, smirnova2009synthesis}
 As expected, the $L_1$ and $L_2$ structures corresponding to FQFE exhibit opposite spin-resolved energy band structures, as shown in Figs. \ref{fig:model}(d) and \ref{fig:model}(e). Based on this model, we can further calculate the polarization associated with FQFE using berry phase method \cite{modernpolar} (also see Supplemental Material \cite{supplementary} Sec. II for details). As shown in Fig. \ref{fig:model}(f), the polarization difference upon the switching process is $\boldsymbol{Q}/2$, where $\boldsymbol{Q}=\frac{e}{\Omega}\times (\boldsymbol{a}+\boldsymbol{b})$ is the polarization quantum corresponding to a 1/2 displacement of the A atom along $\boldsymbol{a}+\boldsymbol{b}$, with $e$ the electron charge and $\Omega$ the volume of the unit cell.

\section{\label{sec:MnTe}$\text{Real system examples--bulk MnTe}$}

\begin{figure}
	\centering
	\includegraphics[width=\linewidth]{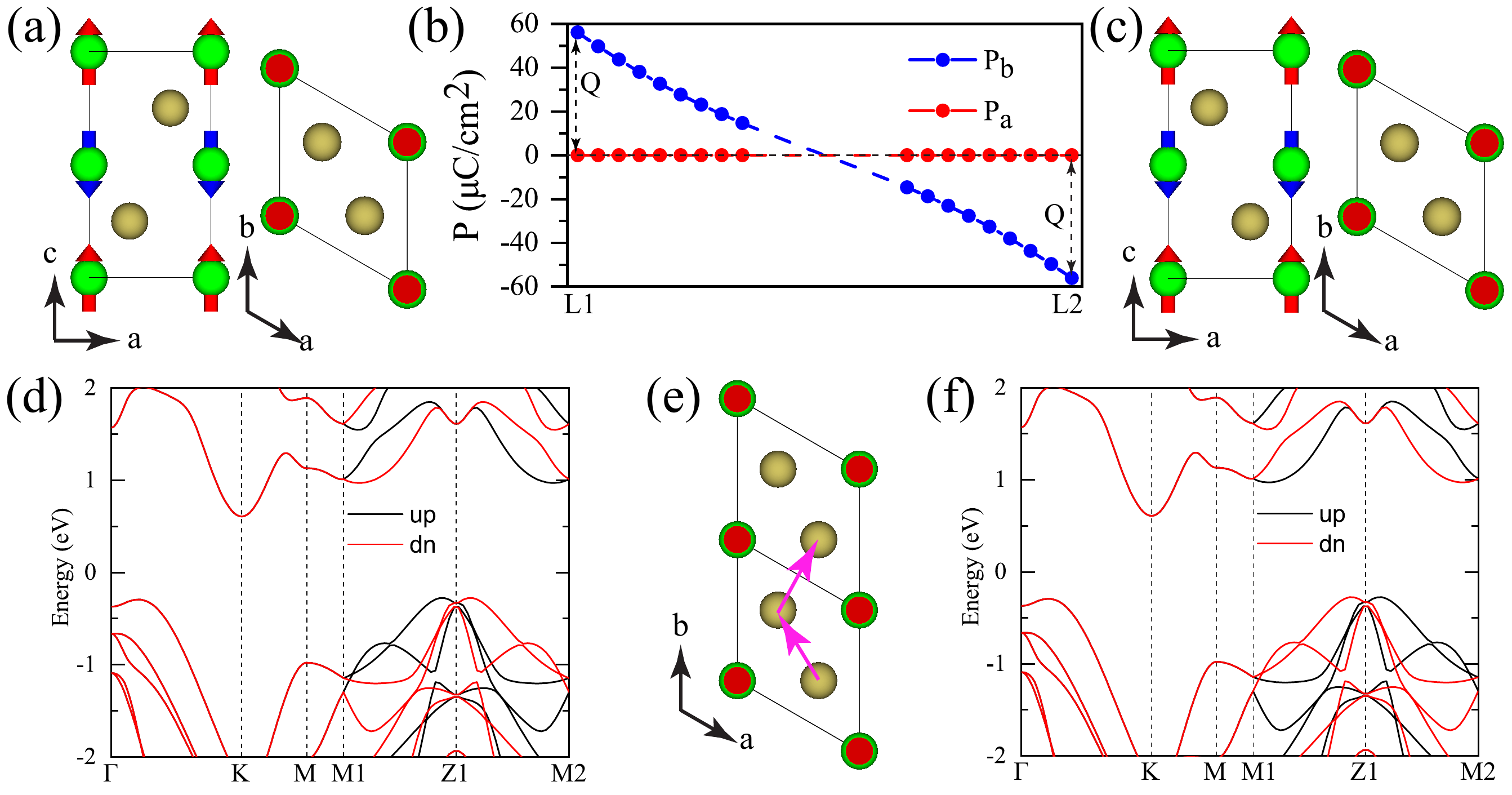}
	\caption{\label{fig:mnte}Schematic illustration, polarization, and band structures of bulk MnTe. (a) $L_1$ and (c) $L_2$ crystal structures of MnTe. The green (yellow) spheres denote Mn (Te) atoms, and arrows indicate spin direction. (b) Evolution of the polarization along the $L_1$--$L_2$ pathway. (d) and (f) Spin-polarized band structure of $L_1$, and $L_2$, respectively. (e) Schematic diagram of atomic motion during the transition from $L_1$ to $L_2$.The pink arrow traces the Te displacement from $L_1$ to $L_2$.}
\end{figure}

We next confirm the existence of FQMF in real materials. As a representative example, we focus on bulk MnTe, which has been experimentally synthesized and identified as an AM material with high Néel temperature ($\sim$ 300 K) \cite{PhysRevX.12.031042, bai2024altermagnetism, krempasky2024altermagnetic}. As shown in Fig. \ref{fig:mnte}(a), MnTe crystallizes in the space group P6$_3$/mmc (No. 194), referred to as the $L_1$ structure. The other FQMF structure, denoted as the $L_2$ structure [Fig. \ref{fig:mnte}(c)], can be generated from $L_1$ structure by applying the $\tau\mathcal{T}$ operation with $\tau=(0, 0, 0.5)$. As illustrated in Fig. \ref{fig:mnte}(e), the $L_1 \rightarrow L_2$ transition displaces two Te atoms: one from (2/3, 1/3, $z_1$) to (1/3, 2/3, $z_1$) and the other from (1/3, 2/3, $z_2$) to (2/3, 4/3, $z_2$), while the Mn atoms remain fixed. Consequently, the net displacement of Te atoms is $\boldsymbol{\Delta d_}{Te}=0\cdot \boldsymbol{a}+\boldsymbol{b}$, where $\boldsymbol{a}$ and $\boldsymbol{b}$ denote the lattice vectors. This feature identifies MnTe as a promising candidate for FQMF, with the polarization difference evaluated as $\boldsymbol{\Delta P}=-2\times \frac{e}{\Omega}\times \boldsymbol{\Delta d_}{Te} =-2\boldsymbol{Q}$, where $\boldsymbol{Q}=\frac{e}{\Omega}\times \boldsymbol{b}$ and Te$^{2-}$ valence state (-2) is used.

We further demonstrate the in-plane polarization induced by FQMF in MnTe using density functional theory (DFT) calculations (see Supplemental Material Sec. III \cite{supplementary} for details). Figure \ref{fig:mnte}(b) displays the evolution of the in-plane polarizations along the $L_1 – L_2$ pathway, obtained via the Berry phase method \cite{modernpolar, PhysRevB.47.1651, resta1992theory, RevModPhys.66.899}. The $L_1$ and $L_2$ structures exhibit opposite in-plane polarizations of -56.106 and 56.106 $\mu C/cm^2$, respectively. The calculated polarization quantum $\boldsymbol{Q}$ for MnTe is 56.106 $\mu C/cm^2$. Thus, the polarization difference between $L_1$ and $L_2$, $\boldsymbol{\Delta P}$, is exactly $2\boldsymbol{Q}$, in excellent agreement with the theoretical analysis above. These results establish bulk MnTe as a promising FQMF material.

Additionally, we confirm that the spin splitting in MnTe can be inverted by reversing the FQFE polarization. The spin-polarized band structure calculations were performed on the $L_1$ and $L_2$ structures, as shown in Figs. \ref{fig:mnte}(d) and \ref{fig:mnte}(f), respectively. The band gap is approximately 0.9 eV, which ensures a suitable insulating background for ferroelectricity. Remarkably, a pronounced spin splitting is observed at valence band maximum (VBM) [M1(-0.5, 0.0, 0.3)$\rightarrow$M2(0.5, 0.0, 0.3), see Supplemental Material Sec. IV \cite{supplementary} for $\mathbf{k}$-points coordinates], with a maximum energy difference of about 0.8 eV. It is noteworthy that the observed splitting is an order of magnitude greater than that in AM and conventional ferroelectrics-based multiferroics \cite{PhysRevX.12.031042, PhysRevX.12.040501, PhysRevX.12.021016}, underscoring the significantly enhanced magnetoelectric coupling in FQMF systems. As shown in Fig. S2, the momentum-dependent splitting $\Delta E_n (\mathbf{k})$ for the VBM exhibits the characteristic features of “g-wave” AM \cite{PhysRevX.12.031042}. Importantly, the $L_1$ and $L_2$ structures possess identical band structures but opposite spin polarization, directly linked by the FQMF switching, thereby demonstrating the nature of FQMF.

\section{\label{sec:cr2s3}$\text{Real system examples--bilayer MnBr}_2$}

\begin{figure}
	\centering
	\includegraphics[width=\linewidth]{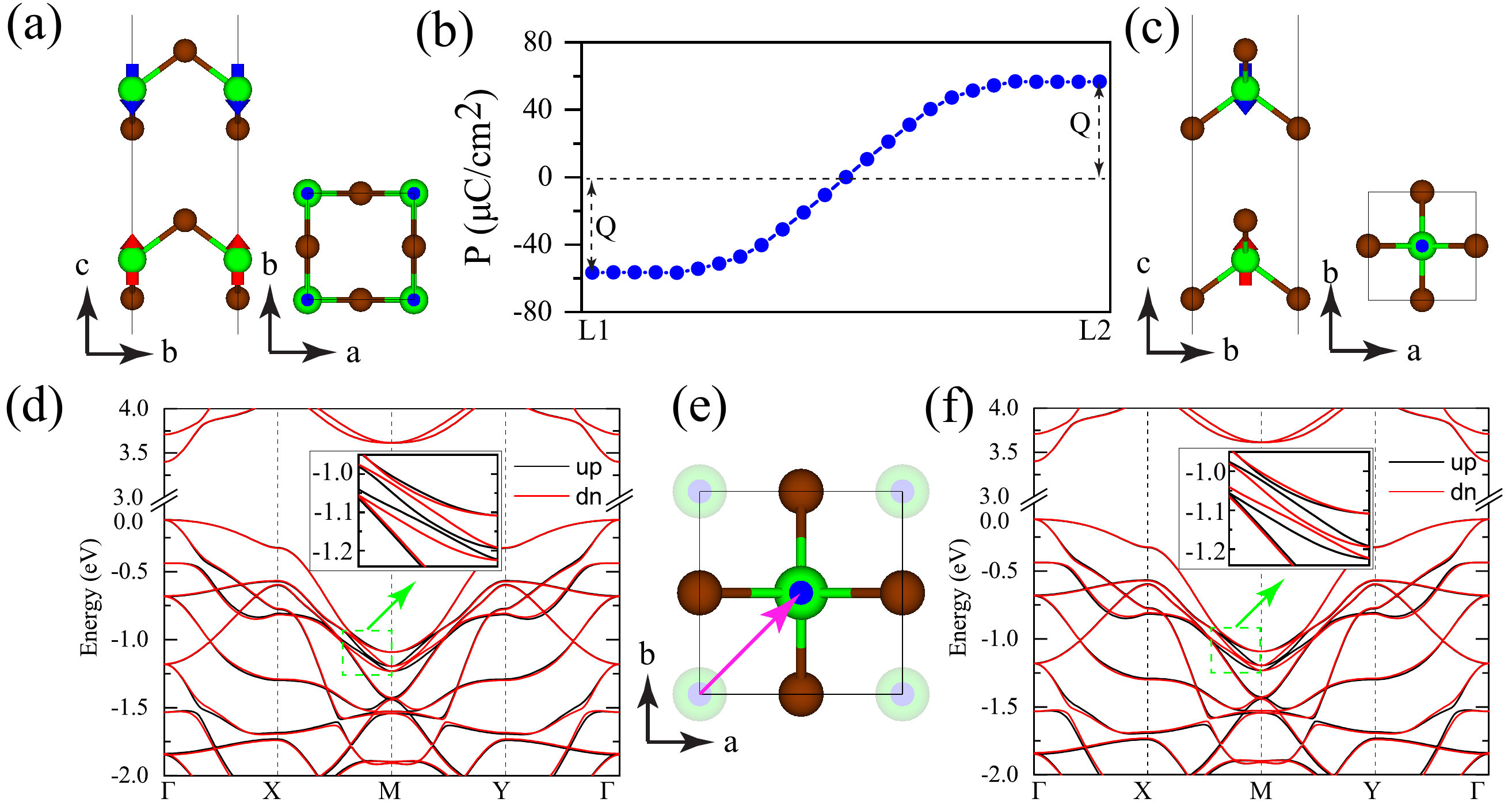}
	\caption{\label{fig:mnbr2}Schematic illustration, polarization, and band structures of bilayer MnBr$_2$. (a) $L_1$ and (c) $L_2$ crystal structures of MnBr$_2$. The green (gray) spheres denote Mn (Br) atoms, and arrows indicate spin direction. (b) Evolution of the polarization along the $L_1$--$L_2$ pathway. (d) and (f) Spin-polarized band structure of $L_1$, and $L_2$, respectively. (e)  Schematic diagram of atomic motion during the transition from $L_1$ to $L_2$. The pink arrow  traces the Mn displacement from $L_1$ to $L_2$.}
\end{figure}

It is worth nothing that the FQMF can also emerge in 2D materials. As an example, we consider bilayer MnBr$_2$, which crystallizes in the space group P$\bar{4}$m2 (No. 115). Figures \ref{fig:mnbr2}(a) and \ref{fig:mnbr2}(c) present the atomic structures of bilayer MnBr$_2$ in the $L_1$ and $L_2$ configurations, respectively. The $L_2$  structure can be obtained from $L_1$ structure via the $\tau\mathcal{PT}$ operation with $\tau = (0.5, 0.5, 0)$. As illustrated in Fig. \ref{fig:mnbr2}(e), during the $L_1 \rightarrow L_2$ transition, Mn atoms in each layer are displaced from (0, 0, $z$) to (0.5, 0.5, $z$), while the Br atoms remain invariant. Consequently, the total displacement of Mn atoms per unit cell is $\boldsymbol{\Delta d_}{Mn}=2\times (\boldsymbol{a}+\boldsymbol{b})$, since each unit cell contains two Mn atoms. Thus, bilayer MnBr$_2$ emerges as a candidate for the FQMF, with polarization difference $\boldsymbol{\Delta P}=2\times \frac{e}{\Omega}\times (\boldsymbol{a}+\boldsymbol{b}) =2\boldsymbol{Q}$ because of the +2 valence state Mn$^{2+}$. We then confirm this in-plane polarization by DFT calculations (see Supplemental Material Sec. III \cite{supplementary} for details). Figure \ref{fig:mnbr2}(b) shows the polarization evolution along the $L_1$--$L_2$ pathway. The two structures possess sizable in-plane polarizations of -56.59 and 56.59 $\mu C/cm^2$, respectively. Moreover, the value of polarization quantum Q is calculated to be 56.59 $\mu C/cm^2$. This result verifies that the polarization difference $\boldsymbol{\Delta P}=2\boldsymbol{Q}$, in agreement with the theoretical analysis above. The spin-resolved band structures of $L_1$ and $L_2$, displayed in Figs. \ref{fig:mnbr2}(d) and \ref{fig:mnbr2}(f), respectively, reveal a large band gap of about 3.52 eV, providing a robust insulating background for ferroelectricity. Notably, the two band structures are identical in energy but exhibit opposite spins, demonstrating that reversing the FQFE polarization inverts $\Delta E_n (\mathbf{k})$. This firmly establishes bilayer MnBr$_2$ as an ideal FQMF material.

\section{\label{sec:tmr}$\text{TMR in MnTe/MnTe FQMFTJ}$}

\begin{figure}
	\centering
	\includegraphics[width=\linewidth]{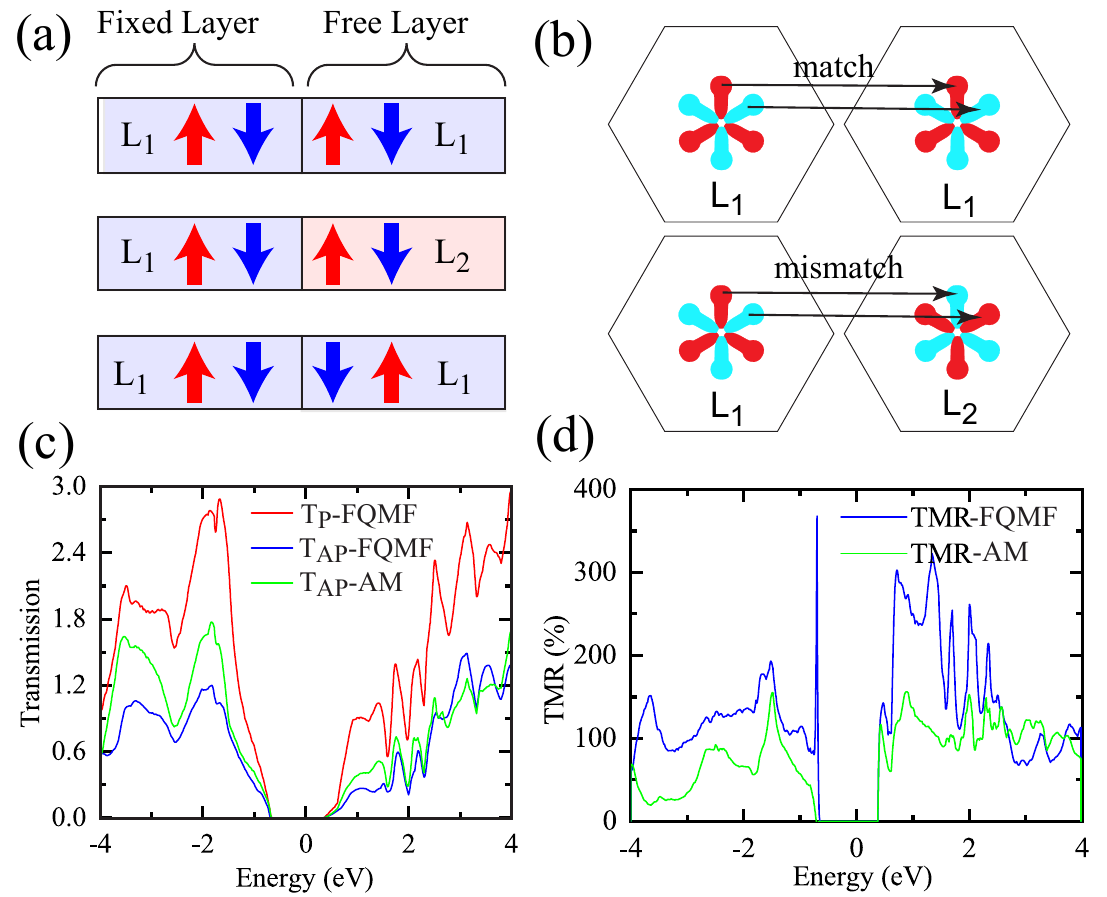}
	\caption{Schematic illustration, Fermi surfaces, transmission spectra, and TMR of FQMFTJ junctions. (a) Schematic illustration of three tunnel junction configurations. In parallel FQMF (P-FQMF), both the fixed and free layers adopt the $L_1$ structure with the same Néel vector. In antiparallel FQMF (AP-FQMF), the fixed layer remains in $L_1$ structure while the free layer transitions to $L_2$ structure, with both layers retaining the same Néel vector. In the antiparallel AM (AP-AM) configuration, both layers are in the $L_1$ structure but possess opposite Néel vectors. (b) Fermi surface slice of MnTe at E= -0.5 eV, $k_z=0.3\pi$, with the red (cyan) representing spin-up (spin-sown) electrons. The top panel shows the P-FQMF ($L_1\rightarrow L_1$) configuration, where spin states are momentum-matched during tunneling. The bottom panel shows the AP-FQMF ($L_1\rightarrow L_2$) configuration, where spin states are momentum-mismatched during tunneling. (c) Energy-resolved transmission spectra for the P-FQMF, AP-FQMF, and AP-AM configurations. (d) Energy-resolved TMR spectra of the FQMFTJ (blue) and AMTJ (green).}
	\label{fig:tmr}
\end{figure}

Finally, we discuss the potential applications of FQMFs in spintronic devices. Specifically, we propose an electrically controllable FQMFTJ, in which both the fixed and free layers are composed of MnTe, as illustrated in Fig. \ref{fig:tmr}(a) [see Fig. S3(a) for simulated atomic structure]. In the parallel FQMF configuration (P-FQMF), both the fixed and free layers adopt the $L_1$ structure with the same Néel vector, leading to identical band structures. Consequently, spin-up (spin-down) electrons in the free layer possess Fermi surfaces identical to those of their spin-up (spin-down) counterparts in the fixed layer, thereby satisfying the momentum-matching condition during tunneling [see Fig. \ref{fig:tmr}(b), top panel]. This results in relatively high transmission and conductivity. In the antiparallel FQMF configuration (AP-FQMF), the fixed layer remains in the $L_1$ structure while the free layer switches to the $L_2$ structure, while both retain the same Néel vector. In this case, the fixed and free layers exhibit identical band structures but with opposite spin polarization. As a result, spin-up (spin-down) electrons in the free layer encounter Fermi surfaces that differ from those in the fixed layer [see Fig. \ref{fig:tmr}(b), bottom panel], breaking the momentum-matching condition and thus suppressing transmission and conductivity. Importantly, the transition from the P-FQMF to the AP-FQMF configuration can be induced by an external electric field.

Figure \ref{fig:tmr}(c) presents the energy-resolved transmission spectra for both the P-FQMF and AP-FQMF configurations, calculated using Quantum-ATK (see Supplemental Material Sec. III \cite{supplementary} for details). Across the energy range from -4.0 to 4.0 eV, the transmission in the P-FQMF configuration consistently exceeds that in the AP-FQMF configuration. Figure \ref{fig:tmr}(d) further shows the energy-resolved Tunnel Magnetoresistance (TMR) ratio, calculated as \cite{ren2022exotic, liu2023electric}
\begin{eqnarray}
	\text{TMR} = 
	\begin{cases}
		\frac{T_P - T_{AP}}{T_{AP}} \times 100\%, & T_{AP} > 0.01 \\
		0, & T_{AP}\leq 0.01
	\end{cases}.
	\label{eq:tmr}
\end{eqnarray}
Note that we only consider TMR with $T_{AP}>0.01$ to avoid numerical divergence caused by excessively small transmission. As shown in Fig. \ref{fig:tmr}(d), the TMR ratio calculated for the FQMFTJ (denoted as TMR-FQMR) exceeds 300\% near the Fermi level, highlighting the potential of FQMFTJs for nonvolatile spintronic applications. It is important to note that in the AP-FQMF configuration, the $L_1$ and $L_2$ structures have different atomic configurations [Fig. S3(a)], which disrupt the periodicity of the atomic lattice at the interface. Consequently, the observed TMR-FQMF arises from two distinct contributions: one from the spin-dependent electronic structure mismatch, similar to AM-based tunnel junctions (AMTJs) \cite{shao2021spin}, and the other from the atomic structure mismatch between the $L_1$ and $L_2$ FQFE structures at the interface, akin to ferroelectric-based tunnel junctions (FTJs) \cite{zhao2020ultrathin, park2024ferroelectric}. Therefore, FQMFTJs inherently outperform tunnel junctions based solely on either AM or ferroelectric materials, due to their integrated characteristics.

We next attempt to isolate the TMR contributions from the AM and FQFE characteristics. To this end, we additionally consider an AMTJ, in which both the fixed and free layers adopt the $L_1$ structure but possess opposite Néel vectors [denoted as AP-AM in Fig. \ref{fig:tmr}(a); note that the P-AM configuration is identical to the P-FQMF]. In this case, the TMR originates solely from the electronic structure mismatch associated with the AM characteristic (see Supplemental Material Sec. VI \cite{supplementary} for a quantitative comparison between TMR-AM and spin-splitting). As shown in Fig. \ref{fig:tmr}(c), the TMR obtained for the AMTJ (TMR-AM) is significantly smaller than TMR-FQMF, owing to the absence of atomic structure mismatch at the interface. This result clearly demonstrates the pronounced advantage of FQMFTJs compared to conventional junctions, where the TMR arises from only a single ferroic order.

Before closing, we want to emphasize that the FQMF exists in a wide range of bulk and 2D materials, including Cr$_2$S$_3$, Mn$_4$Bi$_3$NO$_{15}$, CoCl$_2$, CoBr$_2$, FeI$_2$, MnCl$_2$ and MnI$_2$. To show this, we have performed additionally symmetry analysis on above materials, following the same procedure as above. As a result, it is found that FQFE and AM coexist in these systems. First-principles calculations have also been performed, which confirm that their spin splitting reverses upon switching the ferroelectric polarization, thereby establishing their intrinsic FQMF character (see Supplemental Material Sec. SVII-SIX \cite{supplementary} for details). These results demonstrate that the FQMF has a broad domain of realization.

\section{\label{sec:Conclusion}Conclusion}

In summary, guided by symmetry analysis, we have identified FQMF as a distinct class of multiferroic materials, in which fractional quantum ferroelectric polarization is intrinsically coupled to altermagnetic spin splitting via $\mathcal{PT}$ or $\mathcal{T}\tau$ operations.This mechanism enables the electrical reversal of spin texture without Néel vector reorientation, as demonstrated by a minimal tight-binding model. First-principles calculations further reveal a broad family of two and three dimensional candidate compounds, including bulk MnTe, Cr$_2$S$_3$, Mn$_4$Bi$_3$NO$_{15}$ and 2D bilayers AB$_2$ bilayer MnX$_2$ (X=Cl, Br, I), CoCl$_2$, CoBr$_2$, FeI$_2$. Leveraging MnTe’s high Néel temperature and large electrically switchable spin splitting, we design an electric-field-controlled FQMF tunnel junction with predicted tunneling magnetoresistance exceeding 300\%. These results establish FQMFs as a symmetry-rooted route to strong, room-temperature magnetoelectric coupling and a promising platform for voltage-controlled spintronics.


\section{\label{sec:acknowlodgment}acknowlodgments}
 
This work was supported by the Ministry of Science and Technology of the People’s Republic of China (Grant No. 2022YFA1402901), Natural Science Foundation of China (Grants No. 12474237), Science Fund for Distinguished Young Scholars of Shaanxi Province (Grant No. 2024JC-JCQN-09), and Shanghai Pilot Program for Basic Research—FuDan University 21TQ1400100 (23TQ017).


%

\end{document}